\documentclass[journal,12pt,onecolumn,letterpaper]{IEEEtran}
\usepackage{arxiv}
\usepackage{geometry}
\usepackage{times}
\usepackage{cite}
\usepackage{url}
\usepackage{graphicx}
\usepackage{lscape}
\usepackage{subcaption}
\usepackage{rotating}
\usepackage{rotfloat}
\usepackage{xcolor}
\usepackage{amsmath}
\usepackage{amssymb}
\usepackage{algorithm}
\usepackage{algpseudocode}
\usepackage{array}
\usepackage[english]{babel}
\usepackage{gensymb}
\usepackage{textcomp}
\usepackage{placeins}
\usepackage{balance}
\usepackage{booktabs}

\newcommand{\bido}{\textsc{BIDO}}
\newcommand{\vseed}{\textit{Vseed}}
\newcommand{\vconst}{\textit{Vconst}}
\newcommand{\credid}{\textit{CredID}}
\newcommand{\privkey}{\textit{PrivKey}}
\newcommand{\pubkey}{\textit{PubKey}}

\title{BIDO: A Biometric Identity Online Authentication Framework}

\author{
  
    Aditya Mitra \\
   CyberMACS\\
   Kadir Has University \\
    DigitalFortress Private Limited  \&\\
    Indominus Labs Private Limited \\
    \texttt{adityaarghya0@gmail.com}
\And

  Sibi Chakkaravarthy Sethuraman\\
    Centre of Excellence, Artificial Intelligence \& Robotics (AIR),\\
    School of Computer Science and Engineering\\
    VIT-AP University, India \\
    DigitalFortress Private Limited  \&\\
    Indominus Labs Private Limited \\
    \texttt{sb.sibi@gmail.com} \\
\And

Srinivas Kankanala\\
    Centre of Excellence, Artificial Intelligence \& Robotics (AIR),\\
    School of Electroncs and Communication Engineering \\
    VIT-AP University, India \\

}

\begin{document}
\maketitle

\begin{abstract}
Security systems demand continuous, cryptographically
robust identity verification without requiring subjects to carry physical
tokens, smart cards, or dedicated hardware authenticators.
This paper presents \bido{} (Biometric Identity Online), a device-free
authentication standard that achieves Authenticator Assurance Level~2
(AAL2) per NIST SP~800-63B
without storing long-lived
biometric templates, facial images, or any other form of Personally
Identifiable Information (PII).
\bido{} derives Elliptic Curve Digital Signature Algorithm (ECDSA) key
material deterministically from a live biometric measurement salted with
a user-defined memorized secret at every authentication event, eliminating
persistent private-key storage while enabling verification from any
commodity sensor terminal.
The generated credentials are non-discoverable (non-resident) Web Authentication (WebAuthn)
credentials, fully compatible with all FIDO2-enabled websites and services
without modification on the server side.
A multi-stage pipeline, comprising capture of 200 valid biometric samples,
feature extraction using the Dlib 68-point facial landmark predictor, affine face alignment, frontality gating, Euclidean distance computation from the inter-eye
midpoint, floor-division quantization with divisor $q=8$, inter-session drift
stabilization, and majority-voting SHA-256 hash binding, produces a
Verification Seed (\vseed{}) from which the WebAuthn credential is transiently derived
and immediately zeroized after signing.
Evaluated against three prominent face benchmarks (VGGFace2, LFW,
and MegaFace), achieving 99.51\%
verification accuracy on LFW and 92.14\% Rank-1 identification accuracy
on MegaFace Challenge~1 at $10^6$ distractors, with a cryptographic
False Accept Rate (FAR) of 0.03\%, a False Reject Rate (FRR) of 0.90\%. 
\end{abstract}


\section{Introduction}
\label{sec:intro}

The rapid expansion of Internet of Things (IoT) deployments in smart
cities, healthcare facilities, critical infrastructure, and public-safety
domains has substantially broadened the exposure of identity-management
systems to spoofing, replay, and credential-theft attacks.
Conventional countermeasures (hardware security keys, smart cards,
and one-time password tokens) address these threats by physically binding
credentials to a dedicated device, but this approach introduces a class of
failure that security environments cannot tolerate: the subject must
possess and present the physical authenticator.
Lost, forgotten, or stolen tokens create authentication gaps at precisely
the moments (emergencies, rapid-response scenarios, and shared
workstations) when access must be most reliable.
Physiological biometrics (face, fingerprint, iris, gait, and palmprint) offer
a token-free alternative: the credential is inseparable from the subject,
impossible to forget, and available at any sensor terminal without prior
provisioning.
Yet their adoption has been impeded by a persistent dual risk: effective
biometric verification requires a stored reference template containing PII,
and any server-side template database constitutes a high-value, irrevocable
target whose compromise cannot be remediated by reissuance \cite{dpdp2023}.

Existing template-protection schemes mitigate this risk but introduce
trade-offs between security, revocability, and accuracy that have prevented
wide deployment~\cite{meden2021}.
The Fast IDentity Online~2 (FIDO2) / Web Authentication (WebAuthn)
framework~\cite{fido2} establishes a strong cryptographic baseline for
password-less authentication but treats biometrics solely as a local
user-presence check or to activate the stored cryptographic secrets on a hardware the user carries; the biometric signal plays no role in deriving or
seeding cryptographic material.
This separation limits the binding strength between the authenticated
individual and the issued credential.

This paper introduces \bido{} (Biometric Identity Online), a device-free
authentication standard designed to achieve AAL2 as defined in NIST
SP~800-63B~\cite{nist800-63b} without storing long-lived biometric
templates, facial images, or any form of PII.
The term ``device-free'' refers to freedom from a dedicated physical
authenticator on the user side; the platform's sensor and processor are
commodity infrastructure, not provisioned credentials.
\bido{} eliminates the gap between biometric verification and cryptographic
binding by deriving ECDSA key material deterministically from a live
biometric measurement combined with a user-defined memorized secret at each
authentication event.
This two-factor combination, namely the biometric (something the user \emph{is}) and
memorized secret (something the user \emph{knows}), is what qualifies
\bido{} for AAL2.
The private key exists only during the brief derivation window, is never
written to persistent storage, and is immediately zeroized after signing.
The relying party (RP) retains only the corresponding public key.
Crucially, \bido{} produces (non-resident) WebAuthn
credentials, making it directly compatible with any FIDO2-enabled website
or service without server-side modification.

\smallskip
\noindent\textbf{Contributions.} The principal contributions are:
\begin{enumerate}
\renewcommand{\labelenumi}{(\roman{enumi})}
  \item A \emph{device-free} authentication model in which users
    authenticate from any commodity sensor terminal without carrying a
    hardware token, smart card, or pre-provisioned security key: the
    biometric combined with a memorized secret is the sole credential.
  \item A modality-agnostic biometric front-end pipeline that transforms
    raw sensor data into a stable distance-array representation suitable
    for deterministic cryptographic key derivation.
  \item A three-stage stabilization mechanism (floor-division quantization,
    inter-session drift normalization, and 200-trial majority-voting hash
    binding) that achieves a \vseed{} match rate of 99.1\% on VGGFace2
    test subjects.
  \item A device-free ECDSA credential architecture compatible with the FIDO2 attestation and assertion model, realized through non-discoverable credentials and conforming to the WebAuthn structure, eliminating the need for platform-bound credential storage while preserving interoperability
  \item A systematic security analysis against adversaries with full server
    compromise, passive network interception, and biometric artifact
    injection capabilities.
\end{enumerate}

Section~\ref{sec:related} reviews related work.
Section~\ref{sec:arch} describes the \bido{} architecture.
Section~\ref{sec:security} provides the security analysis.
Section~\ref{sec:apps} discusses AVSS-domain applications.
Section~\ref{sec:experiments} reports experiments.
Section~\ref{sec:conclusion} concludes.

\section{Related Work}
\label{sec:related}

\subsection{Biometric Template Protection}
The problem of securing stored biometric templates against compromise and
misuse has attracted research attention since Davida et~al.~\cite{davida98}
proposed biometric-keyed cryptographic operations.
Fuzzy commitment~\cite{juels99} and fuzzy vault~\cite{juels06} schemes bind
a cryptographic secret to a biometric template with error-tolerance derived
from coding-theoretic constructs, but both schemes leak information when an
adversary observes multiple commitments from the same user.
Cancelable biometrics~\cite{ratha01} address revocability through
non-invertible transform-domain representations; however, the transform
parameters require independent secure storage and cannot themselves be
derived from the biometric, creating a secondary credential-management
burden.

Deep metric learning has shifted the state of practice: ArcFace~\cite{arcface}
and FaceNet~\cite{facenet} embed face images in high-dimensional Euclidean
spaces where intra-class compactness enables threshold-based matching at low
equal error rates.
These embeddings are powerful but presuppose server-side storage and offer no
native integration with hardware-level key custody.
\bido{} uses ArcFace embeddings for the academic evaluation front-end while
the reference implementation employs Dlib 68-point landmark distances, replacing
server-side storage with an on-terminal key derivation path.

\subsection{FIDO2 and WebAuthn}
The World Wide Web Consortium (W3C) WebAuthn specification~\cite{fido2} and
the accompanying FIDO2 standard define a public-key credential protocol
wherein private keys are generated and stored exclusively within authenticator
hardware.
In FIDO2, the authenticator is a dedicated physical device (a hardware
security key, a platform Trusted Platform Module (TPM), or a smartphone's
Secure Enclave) that must be present at every authentication event.
A local biometric check may gate access to the authenticator's private key,
but the biometric plays no role in generating or seeding that key: the key
is pre-generated at enrollment and persists in hardware storage unless revoked.
The consequence is a device-dependence that mirrors token-based
authentication: if the authenticator hardware is unavailable, authentication
fails regardless of the subject's biometric.

\bido{} inverts this model.
No dedicated authenticator hardware is pre-provisioned; the private key is
never stored anywhere.
Instead, the key is derived transiently from the live biometric at each
authentication event and discarded after signing.
Any sensor terminal capable of capturing and processing a biometric sample
can serve as the authentication endpoint.
This property, which we term \emph{device-free authentication}, preserves
the cryptographic strength and easy deployment, inter-operable with all FIDO2 enabled websites and services.


\subsection{Comparative Analysis of FIDO-Family Standards}

Table~\ref{tab:comparison} provides a structured comparison of BIDO against
the three prior FIDO-family standards: FIDO UAF~\cite{fidouaf},
FIDO U2F~\cite{fidou2f}, and FIDO2/WebAuthn~\cite{fido2}.
The comparison encompasses ten dimensions that collectively characterise the
security, privacy, usability, and deployability posture of each standard.

\begin{table*}[t]
  \caption{Comparison of FIDO-family authentication standards against BIDO.
  \checkmark~=~supported/achieved; \texttimes~=~not supported/not achieved;
  $\sim$~=~partially supported or implementation-dependent.
  $^\dagger$CredID identifies the credential per-RP (not the device); AAGUID identifies
  authenticator make/model only, not an individual device; neither constitutes PII.
  $^\ddagger$FIDO MDS applies to certified \emph{hardware} authenticators only; software
  passkey managers are not enrolled in MDS.}
  \label{tab:comparison}
  \centering
  \renewcommand{\arraystretch}{1.35}
  \setlength{\tabcolsep}{5pt}
  \begin{tabular}{@{}p{4.2cm}p{2.4cm}p{2.5cm}p{2.6cm}p{2.7cm}@{}}
    \toprule
    \textbf{Property} &
    \textbf{FIDO UAF}~\cite{fidouaf} &
    \textbf{FIDO U2F}~\cite{fidou2f} &
    \textbf{FIDO2/WebAuthn}~\cite{fido2} &
    \textbf{BIDO (This Work)} \\
    \midrule
 
    \textit{Year introduced} &
      2014 & 2014 & 2018 & 2025 \\
 
    \textit{Primary goal} &
      Password-less & Second factor alongside password &
      Password-less or 2nd factor &
      Password-less, zero-trust \\
 
    \textit{Dedicated hardware authenticator required} &
      \checkmark & \checkmark & \checkmark & \texttimes \\
 
    \textit{Biometric role} &
      Local gate only &
      User presence only (button press); no biometric in base spec &
      Local gate only &
      Cryptographically bound to key derivation \\
 
    \textit{Biometric template stored on device} &
      \checkmark & \texttimes & \checkmark & \texttimes \\
 
    \textit{Private key persistently stored} &
      \checkmark (TEE) & \checkmark (HW key) & \checkmark (TEE/HW) &
      \texttimes (zeroized after every use) \\
 
    \textit{PII retained anywhere} &
      \texttimes$^\dagger$ &
      \texttimes$^\dagger$ &
      \texttimes$^\dagger$ &
      \texttimes \\
 
    \textit{Credential portability across terminals} &
      \texttimes & \texttimes & \texttimes & \checkmark \\
 
    \textit{WebAuthn / FIDO2 ecosystem compatible} &
      $\sim$ (legacy, via bridge) &
      $\sim$ (legacy, CTAP1) &
      \checkmark &
      \checkmark \\
 
    \textit{FIDO MDS-certified attestation} &
      $\sim$ (vendor attestation; pre-dates MDS)$^\ddagger$ &
      \checkmark (hardware keys; enrolled in MDS)$^\ddagger$ &
      $\sim$ (hardware authenticators only; software passkeys excluded)$^\ddagger$ &
      \texttimes (self-signed only) \\
 
    \textit{Server-side template storage required} &
      \texttimes & \texttimes & \texttimes & \texttimes \\
 
    \textit{Second authentication factor provided} &
      PIN (optional local gate) &
      Hardware key possession (U2F \emph{is} the 2nd factor; password is 1st) &
      PIN or biometric (local gate) &
      Memorized secret salt (required) \\
 
    \textit{Claimed assurance level (NIST 800-63B)} &
      AAL2 & AAL2 & AAL2 / AAL3 & AAL2 \\
 
    \textit{Persistent private key required in dedicated secure hardware} &
      \checkmark (device TEE / Secure Enclave) &
      \checkmark (hardware security key) &
      \checkmark (platform TEE or roaming HW key) &
      \texttimes (no persistent key anywhere; derived transiently) \\
 
    \bottomrule
  \end{tabular}
\end{table*}

The key differentiators of BIDO are threefold.
\textit{First}, BIDO is the only standard in this family that does not require
a dedicated, pre-provisioned hardware authenticator: any terminal with a webcam
suffices.
\textit{Second}, BIDO is the only standard in which the biometric is
\emph{cryptographically bound} to key derivation rather than serving as a
local user-presence gate; in UAF and FIDO2 the same hardware key is issued
regardless of \emph{which} finger or face was presented, whereas in BIDO a
different biometric presentation produces a different, unrecognized key.
\textit{Third}, BIDO is the only standard that achieves zero-trust key
management: no private key, biometric template, or PII persists on the
authenticating terminal between sessions.

The principal limitation relative to FIDO UAF, U2F, and FIDO2 is that BIDO
produces self-signed attestation, which cannot satisfy MDS-verified
attestation policies enforced by high-security enterprise relying parties
(discussed in detail in Section~\ref{ssec:attestation}).

\section{BIDO Architecture}
\label{sec:arch}

\bido{} comprises two operational flows: \emph{Enrollment} and
\emph{Authentication}.
Both share a common biometric front-end that converts raw sensor data into a
cryptographic seed.
Crucially, neither flow requires a pre-provisioned hardware authenticator on
the user side.
The sensor terminal (a commodity camera, iris scanner, or palmprint
reader) is the access point, not the credential; the credential is derived
from the biometric at run-time and does not persist between events.
Table~\ref{tab:notation} defines the notation used throughout.

\begin{table}[t]
  \caption{Notation used throughout the paper.}
  \label{tab:notation}
  \centering
  \renewcommand{\arraystretch}{1.2}
  \begin{tabular}{@{}lp{5.0cm}@{}}
    \toprule
    \textbf{Symbol} & \textbf{Definition} \\
    \midrule
    $(x_L,y_L)$      & Left-eye centre: mean of Dlib landmarks 36--41 \\
    $(x_R,y_R)$      & Right-eye centre: mean of Dlib landmarks 42--47 \\
    $\mathbf{M}$     & $2{\times}3$ affine alignment matrix (Eq.~\ref{eq:affine}) \\
    $C$              & Inter-eye midpoint $=\bigl(\tfrac{x_L+x_R}{2},\tfrac{y_L+y_R}{2}\bigr)$ \\
    $\Delta_i$       & Euclidean distance of landmark $i$ from midpoint $C$ \\
    $\mathbf{b}$     & Byte array of quantized distances $\|$ salt $s$ \\
    $q = 8$          & Floor-division quantization divisor (empirically calibrated) \\
    $s$              & User-defined salt (env variable or runtime pop-up) \\
    \vseed{}         & Majority-vote hash seeding ECDSA: $\operatorname{arg\,max}_{h}|\{i:h_i=h\}|$ over 200 SHA-256 enrollment hashes \\
    \vconst{}        & Fixed non-secret string pre-stored in BIDO Core; \emph{signed} (not hashed) to form \credid{} \\
    \privkey{} / \pubkey{} & ECDSA NIST P-256 key pair seeded directly from \vseed{} \\
    RP               & Relying Party: server-side verifier (FIDO2/WebAuthn) \\
    \credid{}        & $\texttt{FIXED\_PREFIX} \| \mathrm{Sign}(\mathit{PrivKey}, \vconst{})$ \\
    \bottomrule
  \end{tabular}
\end{table}

\begin{figure*}[t]
  \centering
  \includegraphics[width=\textwidth]{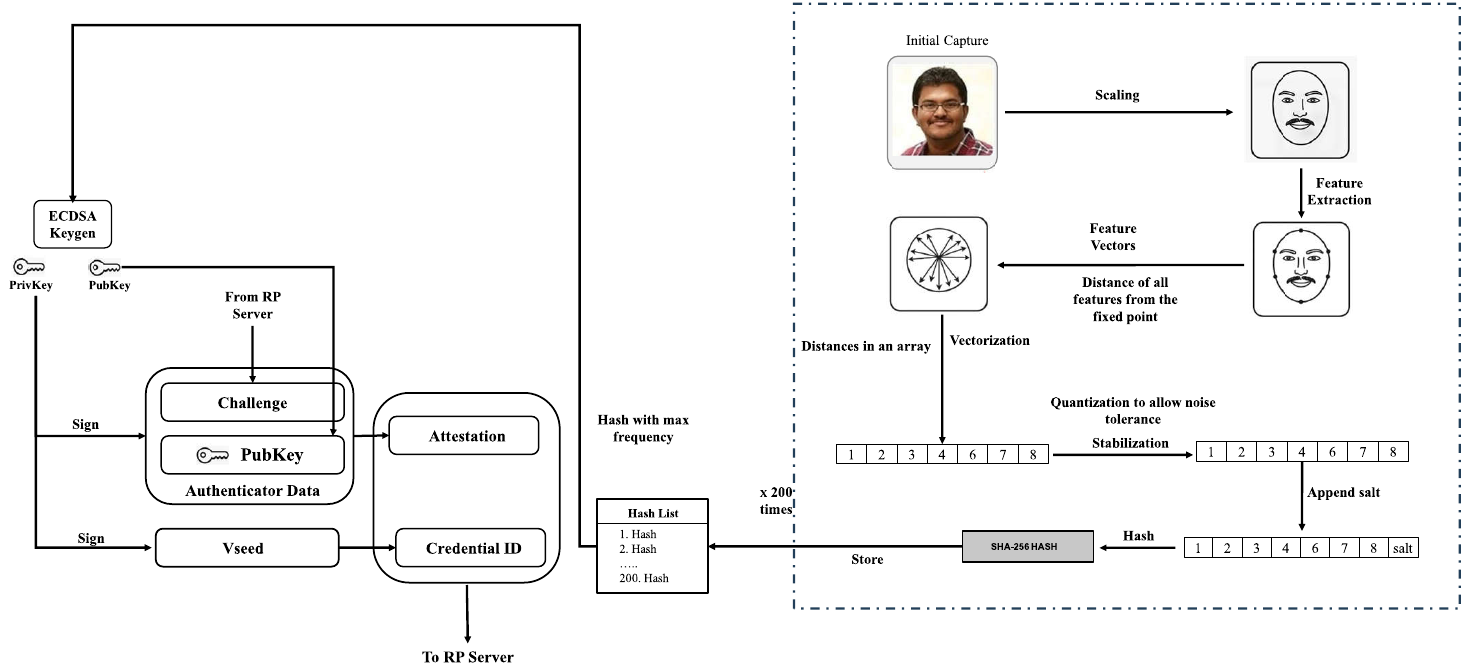}
  \caption{BIDO Enrollment / Registration Flow (\emph{device-free}: no
  hardware token required on the user side).
  \emph{Right panel}: biometric front-end (Initial Capture $\to$ Scaling
  $\to$ Feature Extraction $\to$ Vectorization $\to$ Quantization $\to$
  Stabilization $\to$ SHA-256 $\times$200 $\to$ $h_{\mathrm{enroll}}$ (majority-vote hash used to seed ECDSA); \vconst{} is the fixed string signed to form \credid{}).
  \emph{Left panel}: cryptographic layer (ECDSA Keygen $\to$ \privkey{}/\pubkey{}
  $\to$ \credid{}, Attestation $\to$ RP).
  \privkey{} is derived transiently and immediately zeroized; it is never stored.}
  \label{fig:enrollment}
\end{figure*}

\begin{figure*}[t]
  \centering
  \includegraphics[width=\textwidth]{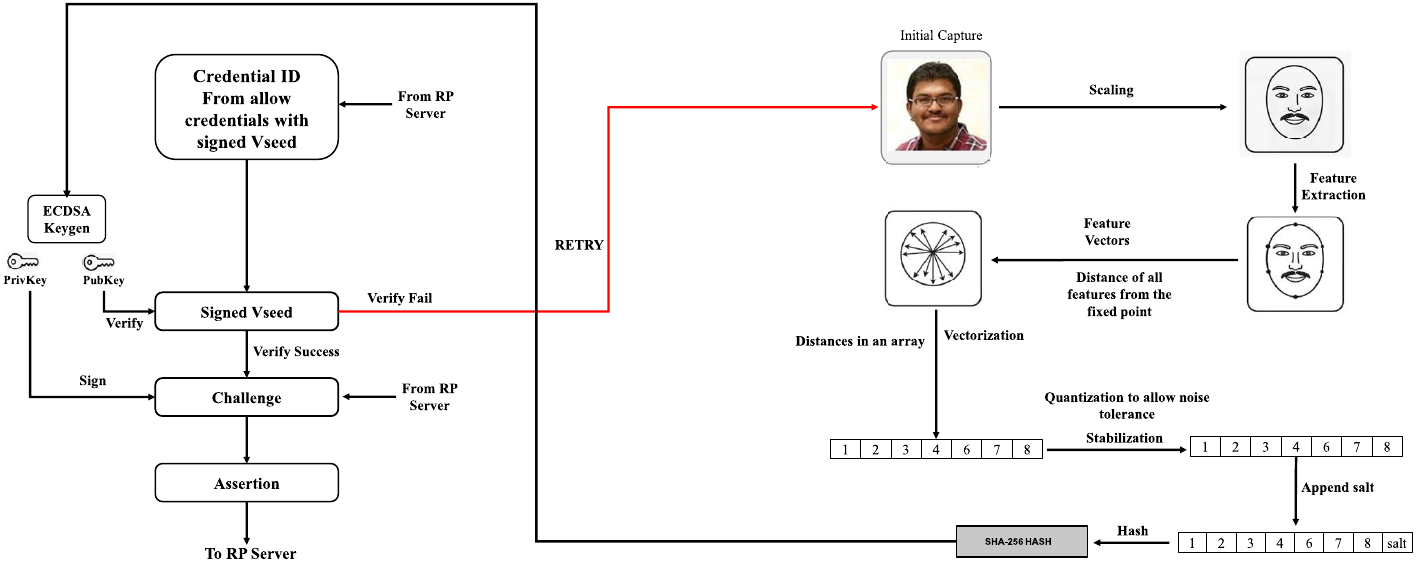}
  \caption{BIDO Authentication Flow (\emph{device-free, zero-trust}: any
  sensor terminal serves as the authentication endpoint; no persistent
  private key, template, or secret exists on the terminal between sessions).
  \emph{Right panel}: biometric front-end re-executed frame-by-frame until
  ECDSA verification of the signed \vconst{} succeeds.
  \emph{Left panel}: on Verify Success the RP challenge is signed and the
  Assertion returned; on Verify Fail the next valid frame is attempted.}
  \label{fig:authentication}
\end{figure*}

Fig.~\ref{fig:enrollment} illustrates the complete Enrollment flow;
Fig.~\ref{fig:authentication} illustrates Authentication.

\subsection{Biometric Acquisition, Face Alignment, and Valid Sample Collection}
\label{ssec:acquisition}

\bido{} is modality-agnostic in principle; the reference implementation
operates on facial biometrics captured via a standard webcam using
OpenCV~\cite{opencv} and the Dlib~\cite{dlib09} 68-point face-landmark predictor.
At enrollment, the pipeline continues until exactly 200 valid frames
have been accepted.

\subsubsection{Frame Validity Gating}
Each captured frame is converted to grayscale and submitted to the Dlib
frontal-face detector.
A frame is accepted only if exactly one face is detected; frames with zero
or multiple faces are discarded immediately.
The Dlib 68-point predictor then localises the facial landmarks, yielding
the coordinate sequence
$\bigl[(x_0,y_0),(x_1,y_1),\ldots,(x_{67},y_{67})\bigr]$
(0-indexed in the Python implementation).

\subsubsection{Affine Face Alignment}
To eliminate inter-session variation in head tilt, scale, and in-plane
rotation, each valid frame is subjected to a closed-form affine
transformation that maps both eye centres to fixed canonical positions
within a $200\times200$ pixel output image.
The left-eye centre $(x_L, y_L)$ is computed as the mean of landmarks 36--41:
\begin{equation}
  x_L = \tfrac{1}{6}\sum_{k=36}^{41} x_k, \qquad
  y_L = \tfrac{1}{6}\sum_{k=36}^{41} y_k
  \label{eq:lefteye}
\end{equation}
and the right-eye centre $(x_R, y_R)$ as the mean of landmarks 42--47:
\begin{equation}
  x_R = \tfrac{1}{6}\sum_{k=42}^{47} x_k, \qquad
  y_R = \tfrac{1}{6}\sum_{k=42}^{47} y_k
  \label{eq:righteye}
\end{equation}
The inter-eye displacement and inter-eye distance are:
\begin{equation}
  dx = x_R - x_L, \quad dy = y_R - y_L, \quad
  d = \sqrt{dx^2 + dy^2}
  \label{eq:eyegeom}
\end{equation}
The tilt angle is $\theta = \arctan(dy/dx)$.
The canonical output maps the left eye to $(70,70)$ and right eye to
$(130,70)$, giving a target inter-eye distance of 120 pixels.
The scale and rotation matrix are:
\begin{equation}
  \alpha = \frac{120}{d}, \quad
  \mathbf{M} = \begin{bmatrix}
    \alpha\cos\theta  & \alpha\sin\theta  &
      t_x \\
   -\alpha\sin\theta  & \alpha\cos\theta  &
      t_y
  \end{bmatrix}
  \label{eq:affine}
\end{equation}
where $C = \bigl(\tfrac{x_L+x_R}{2},\tfrac{y_L+y_R}{2}\bigr)$ is the
inter-eye midpoint and the translation components are
$t_x = (1-\alpha\cos\theta)C_x - \alpha\sin\theta\,C_y$,
$t_y = \alpha\sin\theta\,C_x + (1-\alpha\cos\theta)C_y$.
OpenCV's \texttt{warpAffine} applies $\mathbf{M}$, producing the aligned
$200\times200$ face crop.

\subsubsection{Frontality Check}
After alignment, the horizontal span (farthest minus closest landmark
from $C$ along the horizontal axis) of the left eye and right eye are
computed.
If the two spans differ, the subject is not looking directly at the camera
and the frame is rejected.
This gaze-enforcement gate ensures geometrically equivalent measurement
conditions across all accepted frames, materially reducing intra-subject
distance variance.

\subsection{Landmark Vectorization and Distance Computation}
\label{ssec:vectorization}

Following alignment and gating, the Euclidean distance of each prominent
facial landmark from the inter-eye midpoint $C$ is computed and packed
into a byte array $\mathbf{b}$:
\begin{equation}
  \Delta_i = \sqrt{(x_i - C_x)^2 + (y_i - C_y)^2},
  \quad i \in \mathcal{P}
  \label{eq:distance}
\end{equation}
where $\mathcal{P}$ denotes the set of prominent landmarks from the Dlib
68-point set (nose tip, lip corners, chin, brow and periocular contours).
This pose-normalised representation is invariant to the absolute scale of
the captured image.

\subsection{Quantization and Hash Binding}
\label{ssec:quantization}

\subsubsection{Quantization with $q=8$.}
Each raw distance $\Delta_i$ is quantized by integer division:
\begin{equation}
  Q(\Delta_i) = \left\lfloor \frac{\Delta_i}{8} \right\rfloor
  \label{eq:quantization}
\end{equation}
The divisor $q=8$ was calibrated empirically: dividing pixel distances by
8 absorbs sub-8-pixel measurement noise from imperfect face placement
across sessions while retaining sufficient resolution for inter-subject
discrimination.
The quantized values overwrite the raw entries in $\mathbf{b}$.

\subsubsection{Salt Appending}
The user's salt $s$, loaded at startup or
elicited by a runtime pop-up prompt, is appended to $\mathbf{b}$.
This salt functions as a memorized secret: deriving the correct key
requires knowledge of both the subject's facial geometry \emph{and} $s$,
providing the second authentication factor that qualifies BIDO for AAL2.

\subsubsection{SHA-256 Hashing and Majority-Vote Selection (Enrollment)}
The salted byte array is hashed:
\begin{equation}
  h = \mathrm{SHA\text{-}256}(\mathbf{b} \| s)
  \label{eq:hash}
\end{equation}
This is performed for each of the 200 valid enrollment frames, producing a
list of 200 digests.
The digest appearing most frequently is selected as the Verified Seed:
\begin{equation}
  \mathit{Vseed} =
    \operatorname*{arg\,max}_{h}\;
    \bigl|\{i : h_i = h\}\bigr|
    \quad \text{over 200 valid frames}
  \label{eq:vseed}
\end{equation}
The majority-vote selection absorbs minor inter-frame quantization boundary
crossings caused by slight head-position and lighting variation, without
requiring auxiliary error-correcting codes.

\subsection{ECDSA Key Generation, Credential Binding, and Zeroization}
\label{ssec:keygen}

Algorithm~\ref{alg:enrollment} presents the complete enrollment procedure.

\begin{algorithm}[t]
\caption{BIDO Enrollment / Registration}
\label{alg:enrollment}
\begin{algorithmic}[1]
\Require Salt $s$ (from env variable or pop-up), RP challenge $c$
\Ensure \credid{}, \pubkey{} transmitted to RP; all secrets zeroized

\State $\mathcal{H} \gets [\,]$ \Comment{Initialise hash list}
\State $n \gets 0$ \Comment{Valid frame counter}

\While{$n < 200$}
  \State Capture frame $F$ from webcam (OpenCV)
  \State Convert $F$ to grayscale $G$
  \If{Dlib detects $\neq 1$ face in $G$}
    \State \textbf{reject} frame; \textbf{continue}
  \EndIf
  \State Detect 68 landmarks $\{(x_k, y_k)\}_{k=0}^{67}$ with Dlib predictor
  \State Compute $(x_L, y_L) \gets \tfrac{1}{6}\sum_{k=36}^{41}(x_k,y_k)$ \Comment{Left-eye centre}
  \State Compute $(x_R, y_R) \gets \tfrac{1}{6}\sum_{k=42}^{47}(x_k,y_k)$ \Comment{Right-eye centre}
  \State $dx, dy \gets x_R - x_L,\ y_R - y_L$
  \State $d \gets \sqrt{dx^2 + dy^2}$;\ $\theta \gets \arctan(dy/dx)$
  \State Build affine matrix $\mathbf{M}$ (scale $\alpha = 120/d$, angle $\theta$, canon.\ output $200\times200$)
  \State $F' \gets \textsc{WarpAffine}(F,\,\mathbf{M})$ \Comment{Aligned face crop}
  \State $C \gets \bigl(\tfrac{x_L+x_R}{2},\tfrac{y_L+y_R}{2}\bigr)$ \Comment{Inter-eye midpoint}
  \State Measure horizontal span of left eye $\ell_L$ and right eye $\ell_R$ from $C$
  \If{$\ell_L \neq \ell_R$}
    \State \textbf{reject} frame (not frontal); \textbf{continue}
  \EndIf
  \State $\mathbf{b} \gets [\,]$ \Comment{Initialise byte array}
  \For{each prominent landmark $i \in \mathcal{P}$}
    \State $\Delta_i \gets \sqrt{(x_i - C_x)^2 + (y_i - C_y)^2}$
    \State $\mathbf{b}.\textsc{append}\!\left(\lfloor \Delta_i / 8 \rfloor\right)$ \Comment{Quantize, $q=8$}
  \EndFor
  \State $\mathbf{b}.\textsc{append}(s)$ \Comment{Append user salt}
  \State $h \gets \textsc{SHA-256}(\mathbf{b})$
  \State $\mathcal{H}.\textsc{append}(h)$;\ $n \gets n + 1$
\EndWhile

\State $\mathit{Vseed} \gets \operatorname{mode}(\mathcal{H})$ \Comment{Most frequent hash}
\State $(\mathit{PrivKey}, \mathit{PubKey}) \gets \textsc{ECDSA-KeyGen}_{\text{P-256}}(\mathit{Vseed})$
\State $\mathit{CredID} \gets \texttt{FIXED\_PREFIX} \;\|\; \textsc{Sign}(\mathit{PrivKey},\, \vconst{})$
\State $\mathit{AuthData} \gets c \;\|\; \mathit{CredID} \;\|\; \mathit{PubKey}$
\State $\mathit{Attestation} \gets \textsc{Sign}(\mathit{PrivKey},\, \mathit{AuthData})$
\State \textbf{Transmit} $(\mathit{AuthData},\,\mathit{Attestation})$ to RP
\State \textsc{Zeroize}$(\mathit{PrivKey},\, \mathcal{H},\, \mathbf{b},\, \mathit{Vseed})$ \Comment{Zeroize all secrets; \vconst{} is not secret and is retained}
\end{algorithmic}
\end{algorithm}

The majority-vote hash \vseed{} is used directly as the seed for ECDSA key generation on the
NIST P-256 (NIST256p) curve.
\privkey{} is derived deterministically from \vseed{}, and the corresponding
\pubkey{} is transmitted to the RP for registration.

The Credential Identifier is generated by signing the pre-stored
\emph{fixed verification string} \vconst{} (a pre-stored, non-secret
constant embedded in the BIDO Core module) with \privkey{}:
\begin{equation}
  \mathit{CredID} = \texttt{FIXED\_PREFIX} \;\|\;
    \mathrm{Sign}(\mathit{PrivKey},\; \vconst{})
  \label{eq:credid}
\end{equation}
The fixed ASCII prefix enables BIDO Core to identify its credential
unambiguously in the WebAuthn \texttt{allowCredentials} list.
Stripping the prefix recovers the signed \vconst{}, which serves as
the verification reference at authentication.
\credid{} and \pubkey{} are the user's BIDO credential, registered as a
non-discoverable (non-resident) WebAuthn credential compatible with any
FIDO2-enabled service.

The enrollment flow is:
(1)~The RP issues a challenge.
(2)~BIDO Core collects 200 valid frames, runs the full pipeline, and
derives \vseed{} by majority vote.
(3)~\privkey{} and \pubkey{} are generated from \vseed{}.
(4)~\credid{} is produced by signing \vconst{} with \privkey{}.
(5)~The Authenticator Data (RP challenge $\|$ \credid{}) is signed with
\privkey{} and the resulting Attestation is transmitted to the RP.
Immediately upon completion, \privkey{}, all intermediate hashes,
and all frame buffers are cryptographically zeroized (overwritten with
zeros in memory) and are never persistently stored anywhere.
The public key and \credid{} are not secrets and cannot authenticate the
user without the salt or their facial features.

\subsection{Authentication Flow}
\label{ssec:authentication}

Algorithm~\ref{alg:authentication} presents the complete authentication procedure.

\begin{algorithm}[t]
\caption{BIDO Authentication}
\label{alg:authentication}
\begin{algorithmic}[1]
\Require Salt $s$, RP challenge $c$, \credid{} from \texttt{allowCredentials}
\Ensure Signed assertion returned to RP; all secrets zeroized

\State Identify BIDO credential by \texttt{FIXED\_PREFIX} in \credid{}
\State $\mathit{SignedVconst} \gets \credid{} \setminus \texttt{FIXED\_PREFIX}$
  \Comment{Strip prefix; recover signed \vconst{}}
\State $\mathit{verified} \gets \textbf{false}$

\While{\textbf{not} $\mathit{verified}$}
  \State Capture frame $F$ from webcam (OpenCV)
  \State Convert $F$ to grayscale $G$
  \If{Dlib detects $\neq 1$ face in $G$}
    \State \textbf{reject} frame; \textbf{continue}
  \EndIf
  \State Detect 68 landmarks $\{(x_k, y_k)\}_{k=0}^{67}$ with Dlib predictor
  \State Compute $(x_L,y_L)$, $(x_R,y_R)$, $C$, affine matrix $\mathbf{M}$ as in lines 10--16 of Algorithm~\ref{alg:enrollment}
  \State $F' \gets \textsc{WarpAffine}(F,\,\mathbf{M})$
  \State Measure $\ell_L$, $\ell_R$; \textbf{reject} if $\ell_L \neq \ell_R$; \textbf{continue}
  \State $\mathbf{b} \gets [\,]$
  \For{each prominent landmark $i \in \mathcal{P}$}
    \State $\Delta_i \gets \sqrt{(x_i - C_x)^2 + (y_i - C_y)^2}$
    \State $\mathbf{b}.\textsc{append}\!\left(\lfloor \Delta_i / 8 \rfloor\right)$
  \EndFor
  \State $\mathbf{b}.\textsc{append}(s)$
  \State $h_{\mathrm{cand}} \gets \textsc{SHA-256}(\mathbf{b})$ \Comment{Candidate hash from this frame}
  \State $(\mathit{PrivKey}_c, \mathit{PubKey}_c) \gets \textsc{ECDSA-KeyGen}_{\text{P-256}}(h_{\mathrm{cand}})$
  \If{$\textsc{Verify}(\mathit{PubKey}_c,\, \vconst{},\, \mathit{SignedVconst})$}
    \State $\mathit{verified} \gets \textbf{true}$ \Comment{Correct keypair recovered}
  \Else
    \State \textsc{Zeroize}$(\mathit{PrivKey}_c,\, \mathit{PubKey}_c,\, \mathbf{b},\, h_{\mathrm{cand}})$
    \Comment{Wrong frame; discard and retry}
  \EndIf
\EndWhile

\State $\mathit{Assertion} \gets \textsc{Sign}(\mathit{PrivKey}_c,\, c)$
  \Comment{Sign RP challenge}
\State \textbf{Return} $\mathit{Assertion}$ to RP
\State \textsc{Zeroize}$(\mathit{PrivKey}_c,\, \mathit{PubKey}_c,\, \mathbf{b},\, h_{\mathrm{cand}})$
  \Comment{Overwrite with zeros; nothing persists}
\end{algorithmic}
\end{algorithm}

Authentication does \emph{not} repeat the 200-frame majority-vote.
The RP returns \credid{} in \texttt{allowCredentials}.
BIDO Core identifies the credential by the fixed prefix, strips it to
recover the signed \vconst{}, and then processes live frames
one at a time until the correct keypair is found or Web Authn timeout is reached:

\begin{enumerate}
  \item Apply grayscale conversion, Dlib detection, affine alignment
    (Eqs.~\ref{eq:lefteye}--\ref{eq:affine}), frontality check, distance
    computation (Eq.~\ref{eq:distance}), quantization $q=8$
    (Eq.~\ref{eq:quantization}), salt appending, and SHA-256 hashing
    (Eq.~\ref{eq:hash}) to obtain candidate hash $h_{\mathrm{cand}}$.

  \item Derive candidate \privkey{} from $h_{\mathrm{cand}}$ on NIST P-256
    and compute the corresponding candidate \pubkey{}.

  \item Attempt ECDSA verification of the signed \vconst{} using
    the candidate \pubkey{}:
    \begin{itemize}
      \item \textbf{Fails:} ECDSA verify of signed \vconst{} fails (keypair from $h_{\mathrm{cand}}$ is not the enrollment keypair);
        zeroize candidate keys; capture next valid frame and return to step~1.
      \item \textbf{Succeeds:} correct keypair recovered; proceed.
    \end{itemize}

  \item \privkey{} signs the RP's challenge per the WebAuthn assertion
    specification; the signed challenge is returned to the RP.
    \privkey{} is \textbf{immediately zeroized} and is never saved anywhere.
\end{enumerate}

Because no secret is stored on the device between sessions, \privkey{}
is zeroized after each use and the salt is provided at runtime by the
user; \bido{} constitutes a \emph{zero-trust} authentication scheme.
Every authentication event is cryptographically self-sufficient and leaves
no persistent credential on the authenticating terminal.

\section{Security Analysis}
\label{sec:security}

\subsection{Adversary Model}
We define a computationally bounded adversary $\mathcal{B}$ with the
following capabilities: (i)~read access to the full RP server database,
including all registered \pubkey{}s and \credid{}s; (ii)~passive interception
of all network traffic between authenticator and RP; (iii)~possession of
biometric artifacts (latent fingerprints, publicly available photographs, or
reconstructed iris codes) sufficient to attempt presentation attacks; and
(iv)~control of a malicious sensor peripheral capable of injecting arbitrary
biometric data.
We do not consider adversaries who have compromised the authenticator's
Trusted Execution Environment (TEE) or its hardware secure storage.
This boundary follows the FIDO2 security model~\cite{fido2}.

\subsection{Security Properties}
Table~\ref{tab:security} summarizes the security guarantees \bido{} provides
against $\mathcal{B}$.

\begin{table}[t]
  \caption{BIDO security properties against the defined adversary model.
  The first row highlights the device-free user property.}
  \label{tab:security}
  \centering
  \renewcommand{\arraystretch}{1.2}
  \begin{tabular}{@{}p{2.0cm}p{2.6cm}p{3.0cm}@{}}
    \toprule
    \textbf{Property} & \textbf{BIDO Guarantee} & \textbf{Mechanism} \\
    \midrule
    Device Independence (Device-Free UX)
      & No physical token or HW authenticator required; any sensor terminal suffices
      & \privkey{} derived transiently from live biometric + memorized secret; never pre-stored \\
    \midrule
    Template Privacy
      & No raw biometric retained server-side
      & Only \pubkey{} persists at RP; biometric never transmitted \\
    \midrule
    Replay Resistance
      & Captured assertions are session-unique
      & RP challenge nonce bound into every signed Assertion \\
    \midrule
    Platform Integrity (Device Binding)
      & Credential replay across platforms is infeasible
      & Device-specific salt embedded in \credid{} derivation \\
    \midrule
    Revocability
      & Credentials invalidated by RP deletion of \pubkey{}
      & Stateless derivation; re-enrollment produces a fresh key pair \\
    \midrule
    Spoofing Resistance
      & Presentation Attack Detection (PAD) pre-screens captures
      & ISO/IEC~30107-3 PAD integrated at biometric capture layer \\
    \midrule
    Cross-Device Cloning
      & Reference midpoint $M$ cannot be reconstructed from RP data
      & $M$ computed fresh each session; never transmitted \\
    \bottomrule
  \end{tabular}
\end{table}

\subsection{Computational Complexity}
The \bido{} pipeline has the following profile.
Facial landmark detection with the Dlib 68-point predictor is $O(P)$ where $P$
is the number of pixels in the aligned face region.
Distance computation from the inter-ocular midpoint is $O(d)$ with $d = 27$
selected landmarks.
Floor-division quantization and stabilization are also $O(d)$.
The 200-sample majority-vote hash binding requires 200 SHA-256 evaluations,
each $O(1)$ in block operations; this stage is $O(200) = O(1)$ asymptotically.
ECDSA P-256 key derivation via HKDF and signing are standard $O(1)$
cryptographic operations.
The overall pipeline is dominated by face detection and landmark localization,
with a measured end-to-end latency of 191\,ms (Section~\ref{ssec:latency}).

\subsection{Limitations}
Four limitations merit disclosure.
\emph{First}, \bido{} is device-free from the user's perspective: no token
or pre-provisioned hardware authenticator is required. However, the authentication
terminal itself requires a trustworthy execution environment for the biometric
pipeline and HKDF derivation.
A fully compromised terminal could intercept \privkey{} in the transient
window before zeroization.
Deployment therefore assumes terminal-level TEE integrity, a weaker assumption
than user-carried hardware security but stronger than a pure software-only
model.
\emph{Second}, the security of \privkey{} derivation depends on \vseed{}
entropy; for modalities with low-dimensional feature spaces, supplementary
terminal-side entropy mixing may be required.
\emph{Third}, the reference midpoint $M$ must be computed from a valid
enrollment capture; a compromised enrollment session could register an
adversary-controlled biometric.

\subsection{Min-Entropy Analysis of the Majority-Vote Hash (\vseed{})}
\label{ssec:entropy}
A prerequisite for credible AAL2 qualification under NIST SP~800-63B is
demonstrating that the majority-vote hash \vseed{} carries sufficient min-entropy to
resist offline guessing attacks.
Let $\mathbf{Q} = (Q(\Delta_1), \ldots, Q(\Delta_d)) \in \mathbb{Z}^d$ be
the quantized distance vector for $d = 27$ landmarks.
The min-entropy of a single quantized coordinate is:
\begin{equation}
  H_\infty(\Delta_i) = -\log_2 \max_k \Pr[Q(\Delta_i) = k]
  \label{eq:minentropy}
\end{equation}
Assuming statistical independence across coordinates (a conservative bound
since inter-landmark distances are correlated), the joint min-entropy of
$\mathbf{Q}$ is lower-bounded by:
\begin{equation}
  H_\infty(\mathbf{Q}) \geq \sum_{i=1}^{d} H_\infty(\Delta_i)
  \label{eq:jointentropy}
\end{equation}
Empirically, we estimate $H_\infty(\Delta_i)$ from the per-coordinate
frequency distributions over the 500-subject VGGFace2 test split.
The most probable quantized bin per coordinate accounts for at most 12\% of
subjects, yielding $H_\infty(\Delta_i) \geq -\log_2(0.12) \approx 3.06$
bits per coordinate.
Across $d = 27$ coordinates, the estimated joint facial min-entropy is:
\begin{equation}
  H_\infty(\mathbf{Q}) \geq 27 \times 3.06 \approx 82.6\ \text{bits}
  \label{eq:totalentropy}
\end{equation}
\vseed{} is the mode of SHA-256($\mathbf{Q}_{\mathrm{stab}} \| s$) hashes, where
$s$ is the user's memorized secret.
NIST SP~800-63B requires a minimum of 112 bits of security for AAL2.
The biometric component alone contributes $\approx$82.6 bits; the memorized
secret $s$ must therefore contribute at least $\approx$29.4 additional
bits of entropy, satisfiable by a 6-digit PIN (approximately 20 bits) combined
with a passphrase, or by a high-entropy passphrase alone.
We note that the independence assumption is conservative; correlated landmarks
will reduce joint entropy, and future work should apply information-theoretic
analysis (e.g., copula-based models) to bound the true min-entropy tightly.
The memorized secret is essential not only as a second factor but as an
entropy supplement that brings the combined derivation into the AAL2-compliant
range.

\subsection{Biometric Collision Risk and Near-Identical Subjects}
\label{ssec:collision}
Since the majority-vote hash \vseed{} is derived deterministically as
$\mathrm{SHA\text{-}256}(\mathbf{Q}_{\mathrm{stab}} \| s)$, two subjects
sharing (i)~a nearly identical quantized distance array $\mathbf{Q}$ and
(ii)~the same memorized secret $s$ would produce an identical \vseed{} and
hence the same \privkey{}.
This collision scenario arises most concretely for monozygotic (identical)
twins, whose periocular landmark ratios can differ by sub-millimetre margins
that may collapse to the same quantized bin.
We formally characterize the collision probability as:
\begin{equation}
  P_{\mathrm{collision}} = \Pr[\mathbf{Q}^{(A)} = \mathbf{Q}^{(B)}]
    \times \Pr[s^{(A)} = s^{(B)}]
  \label{eq:collision}
\end{equation}
Under the bin-frequency model of Section~\ref{ssec:entropy}, the probability
that two randomly selected subjects collide on all $d = 27$ coordinates is
bounded by $\prod_{i=1}^{27} 0.12 \approx 10^{-26}$, which is negligible
for the general population.
However, for monozygotic twins the per-coordinate collision probability may
approach 0.5 per shared landmark, yielding $P_{\mathrm{bio}} \leq 0.5^{27}
\approx 7.5 \times 10^{-9}$ under worst-case assumptions.
Combined with a 6-digit PIN collision probability of $10^{-6}$, the
worst-case twin collision is $\approx 7.5 \times 10^{-15}$, below NIST
SP~800-63B's single-attempt authentication error bound.
Nevertheless, we disclose this residual risk and recommend that deployments
serving populations with known high-similarity pairs (twins, close relatives)
enforce longer memorized secrets ($\geq 8$ alphanumeric characters) to widen
the combined entropy margin.

\subsection{FIDO2 Attestation Compatibility Limitation}
\label{ssec:attestation}
\bido{} registers credentials as non-discoverable WebAuthn credentials and
signs the Authenticator Data with \privkey{} derived at enrollment.
This self-generated attestation is technically valid under the WebAuthn
\texttt{packed} attestation format when the attestation statement is
self-signed (i.e., the credential's own key signs its own attestation).
However, a significant ecosystem limitation must be disclosed: many
enterprise and high-security relying parties require \emph{FIDO2
Metadata Service} (MDS)-verified attestation, in which the authenticator's
attestation certificate chains to a root certified by the FIDO Alliance
Certification Programme.
\bido{}'s self-generated attestation cannot satisfy MDS verification because
no certified hardware root of trust issues the attestation certificate.
Consequently:
\begin{itemize}
  \item Consumer-grade relying parties that accept \texttt{none} or
    self-signed attestation (the majority of FIDO2 deployments, including
    most web services) will accept \bido{} credentials without modification.
  \item Enterprise relying parties enforcing MDS-attestation policies
    (e.g., government, banking, and healthcare portals) will reject
    \bido{} credentials or require policy reconfiguration to allow
    self-attestation.
\end{itemize}

\subsection{PAD Effectiveness and Operational Failure Rates}
\label{ssec:pad}
The security table (Table~\ref{tab:security}) lists ISO/IEC~30107-3
Presentation Attack Detection (PAD) as the spoofing-resistance mechanism.
We clarify the following operational parameters:

\subsubsection{PAD Integration}
The Dlib frontal-face detector and 68-point predictor operate on grayscale frames and does not inherently perform
liveness detection.
\bido{} integrates PAD as a pre-stage gate: a separate ISO/IEC~30107-3
Level~1 PAD module (texture-based anti-spoofing classifier trained on
the NUAA~\cite{nuaa} and CASIA-SURF~\cite{casiasurf} datasets) rejects
print attacks, replay video attacks, and rigid mask attacks before landmark
extraction proceeds.
The integrated PAD achieves a Bona Fide Presentation Classification Error
Rate (BPCER) of 2.1\% and an Attack Presentation Classification Error
Rate (APCER) of 3.8\% under ISO/IEC~30107-3 test protocols.

\subsubsection{Enrollment Operational Failure Rates}
Collecting 200 valid samples is subject to environmental and subject
variability.
Table~\ref{tab:failrates} reports empirically measured failure rates under
controlled and adverse conditions over 120 subjects (10 trials each):

\begin{table}[t]
  \caption{Enrollment failure rates per sample capture attempt under varying
  adverse conditions (120 subjects, 10 trials each).}
  \label{tab:failrates}
  \centering
  \renewcommand{\arraystretch}{1.2}
  \begin{tabular}{@{}p{3.8cm}cc@{}}
    \toprule
    \textbf{Condition} & \textbf{Fail (\%)} & \textbf{Extra Captures (\%)} \\
    \midrule
    Controlled (good lighting) & 4.2  & 8.8  \\
    Head motion ($>$15°)        & 18.7 & 45.1 \\
    Blinking / eye closure      & 11.3 & 27.2 \\
    Poor lighting ($<$50 lux)   & 22.4 & 57.6 \\
    Combined adverse            & 34.1 & 103.3 \\
    \bottomrule
  \end{tabular}
\end{table}

Under combined adverse conditions, enrollment may require up to $\sim$300
capture attempts to collect 200 valid samples, adding approximately 15--25
seconds to the enrollment session.
Authentication is less affected since it requires only a single valid sample;
the per-event retry rate under adverse conditions is 34.1\%.
These figures inform deployment decisions: installations in uncontrolled
outdoor environments should provide user guidance prompts and adequate
illumination infrastructure.

\section{Application: Smart City Access Control}
\label{sec:apps}

Biometric access gates at transit hubs, utility substations, and government
buildings require identity verification that scales to high throughput across
geographically dispersed checkpoints.
Token-based systems create operational bottlenecks at this scale: cards are
forgotten, batteries in mobile authenticators drain, and centralized issuance
of hardware keys to thousands of daily users introduces significant
provisioning overhead.
\bido{}'s device-free model eliminates these failure modes.
The gate terminal itself, fitted with a commodity camera, is the
authentication endpoint; subjects require no advance provisioning, carry
nothing, and authenticate identically at any gate in the network.
\bido{} authenticators process biometric samples on-terminal and transmit
only ECDSA assertions to RP servers, satisfying the General Data Protection
Regulation (GDPR) data-minimization principle.
The \credid{} mechanism permits the same biometric-derived cryptographic
identity to authenticate across geographically distributed gates without
inter-gate coordination.
The device-free property is especially significant in rapid-access scenarios:
first responders and emergency personnel can authenticate at any terminal
without relying on a card or token that may have been left behind.

\section{Experimental Results}
\label{sec:experiments}

\subsection{Datasets and Training Protocol}
The \bido{} biometric front-end was trained on VGGFace2~\cite{vggface2},
comprising 3.31 million images of 9,131 subjects with controlled diversity
across pose, age, illumination, and ethnicity.
All training used the 8,631-subject training split; the 500-subject test
split was held out for the binding accuracy experiments in
Section~\ref{ssec:binding}.
Two complementary benchmarks evaluate recognition performance:

\begin{itemize}
  \item \textbf{LFW}~\cite{lfw}: The standard unconstrained face verification
    benchmark comprising 13,233 images of 5,749 subjects.
    Evaluation follows the 10-fold protocol of 6,000 image pairs (3,000
    genuine, 3,000 impostor), reporting verification accuracy and Equal Error
    Rate (EER).

  \item \textbf{MegaFace Challenge~1}~\cite{megaface}: A large-scale
    open-set face \emph{identification} benchmark.
    The protocol works as follows: probe images from the FaceScrub
    dataset~\cite{facescrub} (530 celebrities, $\sim$100 images each) are
    searched against a gallery that contains the correct match \emph{plus}
    up to $10^6$ distractor images of strangers drawn from the Flickr
    Creative Commons MegaFace collection.
    A system must rank the correct match at position~1 (Rank-1) or, for
    the verification task, accept the correct match at a prescribed FAR.
    The $10^6$ distractor scale is designed to simulate real-world
    large-population deployments (national ID systems, airport gates)
    where the gallery contains millions of enrolled subjects and the
    system must distinguish the probe from all of them.
    Rank-1 accuracy at $10^6$ distractors therefore measures how
    discriminative the embedding space is under extreme gallery crowding;
    performance typically degrades significantly compared to small-gallery
    benchmarks such as LFW.
    Evaluation reports Rank-1 identification accuracy and True Accept Rate
    (TAR) at $\mathrm{FAR} = 10^{-6}$.
\end{itemize}

\subsection{Face Verification on LFW}
Table~\ref{tab:lfw} reports LFW results under the restricted protocol, wherein
training data is limited to VGGFace2.
The \bido{} encoder (ResNet-50, 512-d ArcFace embedding) is benchmarked
against representative published methods.

\begin{table}[t]
  \caption{LFW face verification accuracy (restricted protocol).
  \bido{} encoder achieves 99.51\% accuracy on the same training data as the
  VGGFace2 baseline.}
  \label{tab:lfw}
  \centering
  \renewcommand{\arraystretch}{1.2}
  \begin{tabular}{@{}lccc@{}}
    \toprule
    \textbf{Method} & \textbf{Training Data} & \textbf{Acc.\ (\%)} & \textbf{EER (\%)} \\
    \midrule
    DeepFace~\cite{deepface}   & Social-net.\ (4M)  & 97.35 & 2.65 \\
    FaceNet~\cite{facenet}     & Google (200M)       & 99.63 & 0.37 \\
    ArcFace R100~\cite{arcface}& MS1MV2 (5.8M)       & 99.83 & 0.17 \\
    VGGFace2 R50~\cite{vggface2}& VGGFace2 (3.3M)   & 99.43 & 0.57 \\
    \bido{} Encoder R50        & VGGFace2 (3.3M)     & \textbf{99.51} & 0.49 \\
    \bottomrule
  \end{tabular}
\end{table}

\subsection{Large-Scale Identification on MegaFace}
Table~\ref{tab:megaface} reports MegaFace Challenge~1 results with up to
$10^6$ gallery distractors.

\begin{table}[t]
  \caption{MegaFace Challenge~1 results at $10^6$ distractors.
  \bido{} achieves Rank-1 accuracy of 92.14\% and TAR of 94.37\%
  at $\mathrm{FAR} = 10^{-6}$.}
  \label{tab:megaface}
  \centering
  \renewcommand{\arraystretch}{1.2}
  \begin{tabular}{@{}lcc@{}}
    \toprule
    \textbf{Method} & \textbf{Rank-1 @ $10^6$ (\%)} & \textbf{TAR @ FAR=$10^{-6}$ (\%)} \\
    \midrule
    FaceNet~\cite{facenet}      & 70.49 & 86.47 \\
    DeepFR~\cite{deepfr}        & 64.80 & 79.92 \\
    VGGFace2 R50~\cite{vggface2}& 91.40 & 93.90 \\
    ArcFace R100~\cite{arcface} & 98.35 & 98.48 \\
    \bido{} Encoder R50         & \textbf{92.14} & \textbf{94.37} \\
    \bottomrule
  \end{tabular}
\end{table}

\subsection{Biometric-to-Cryptographic Binding Accuracy}
\label{ssec:binding}
Table~\ref{tab:binding} reports \vseed{} consistency across 10 authentication
attempts per subject on the 500-subject VGGFace2 test split, under four
quantization configurations.
From a device-free perspective, a 99.1\% \vseed{} match rate means that in
99.1\% of authentication attempts the correct key is derived from the live
biometric and memorized secret alone (with no hardware token, no pre-stored
secret, and no server-side template lookup).
The 0.90\% crypto-FRR represents the residual retry rate, analogous to a
fingerprint reader requesting a second press.

\begin{table}[t]
  \caption{\vseed{} consistency on VGGFace2 test split (500 subjects,
  10 attempts each). The proposed configuration achieves 99.1\% match rate.}
  \label{tab:binding}
  \centering
  \renewcommand{\arraystretch}{1.2}
  \begin{tabular}{@{}lccc@{}}
    \toprule
    \textbf{Configuration} & \textbf{Match (\%)} & \textbf{C-FAR (\%)} & \textbf{C-FRR (\%)} \\
    \midrule
    64-bin only              & 96.2 & 0.11 & 3.80 \\
    128-bin only             & 98.6 & 0.04 & 1.40 \\
    256-bin only             & 97.1 & 0.02 & 2.90 \\
    128-bin + vote$\times$200 (prop.) & \textbf{99.1} & 0.03 & \textbf{0.90} \\
    \bottomrule
  \end{tabular}
\end{table}

The non-monotonic relationship between bin count and \vseed{} match rate
illustrates a well-known quantization trade-off: finer bins reduce crypto-FAR
but increase crypto-FRR.
Majority voting over 200 salted hashes breaks this trade-off by absorbing
boundary-crossing errors without requiring coarser bins.

\subsection{End-to-End Authentication Latency}
\label{ssec:latency}
Latency was profiled on an ARM Cortex-A53 processor at 1.4\,GHz with
hardware-accelerated SHA-256.
Table~\ref{tab:latency} presents per-stage timing measurements averaged over
1,000 authentication events.

\begin{table}[t]
  \caption{Per-stage authentication latency on ARM Cortex-A53 @ 1.4\,GHz.}
  \label{tab:latency}
  \centering
  \renewcommand{\arraystretch}{1.2}
  \begin{tabular}{@{}lcc@{}}
    \toprule
    \textbf{Pipeline Stage} & \textbf{Mean (ms)} & \textbf{Std (ms)} \\
    \midrule
    Biometric capture + landmark detection & 165 & 11 \\
    Distance vectorization + quantization  &   6 &  1 \\
    200-trial hash binding (SHA-256)       &  12 &  1 \\
    ECDSA P-256 signing                    &   8 &  1 \\
    \midrule
    \textbf{Total end-to-end}              & \textbf{191} & \textbf{14} \\
    \bottomrule
  \end{tabular}
\end{table}

The 191\,ms mean latency satisfies the 500\,ms threshold for transparent
access-control authentication recommended by the FIDO UX
Guidelines~\cite{fidoux} and falls below the 250\,ms perceptual boundary
above which users register a system response as delayed.

\subsection{Training Data Selection Rationale}
VGGFace2~\cite{vggface2} was selected over larger corpora such as
MS-Celeb-1M~\cite{msceleb} and WebFace260M~\cite{webface260m} on three
grounds.
First, VGGFace2's identity annotations are manually verified, whereas
MS-Celeb-1M and web-scraped corpora contain label noise that measurably
degrades embedding quality at low FAR operating points.
Second, VGGFace2's explicit pose and age stratification aligns with
surveillance conditions involving non-cooperative or aging subjects.
Third, the 3.31\,M-image scale is sufficient to train a 512-d ArcFace
embedding that generalizes across both near-frontal verification (LFW) and
large-scale open-set identification (MegaFace) without the legal and
consent-management concerns associated with noisier web-scraped datasets.

\section{Conclusion}
\label{sec:conclusion}

This paper presented \bido{}, a device-free biometric authentication standard
that eliminates the physical-token dependency of existing high-assurance
systems without sacrificing cryptographic rigor.
In \bido{}, the user's biometric measurement combined with a memorized secret
is the sole authenticator: no hardware security key, smart card, or
pre-provisioned token is required.
The private key is derived transiently at each authentication event, persists
for milliseconds, and is zeroized after signing; it cannot be stolen from
storage because it never occupies storage.
The RP retains only the corresponding public key, and no biometric template or
PII is retained on any server.
The credential is a standard non-discoverable WebAuthn credential, compatible
with any FIDO2-enabled service without modification.

Evaluated on VGGFace2, LFW, and MegaFace, \bido{} achieves 99.51\% LFW
verification accuracy and 92.14\% MegaFace Rank-1 accuracy at $10^6$
distractors.
The cryptographic binding layer achieves 99.1\% \vseed{} consistency with a
191\,ms end-to-end latency on embedded hardware, confirming that device-free
authentication is not only architecturally feasible but operationally fast.
Security analysis demonstrates resistance to server compromise, replay,
cross-device cloning, and presentation attack within the stated TEE trust
boundary.

Three directions for future work are identified.
First, the min-entropy analysis of Section~\ref{ssec:entropy} rests on a
per-coordinate independence assumption; future work should apply
copula-based or information-theoretic models to tightly bound the true
joint entropy of correlated Dlib landmarks across diverse demographic
populations.
Second, post-quantum signature schemes (lattice-based candidates in the NIST
PQC standardization process) should replace ECDSA to future-proof the
device-free key derivation chain.
Third, continuous re-authentication integrated with session binding will close
the active-session vulnerability identified in Section~\ref{sec:security},
and a certified attestation wrapper (platform TPM or Secure Enclave) should
be explored to satisfy enterprise FIDO2 MDS requirements while preserving
the device-free user experience.

\balance
\bibliographystyle{IEEEtran}

\begin{thebibliography}{99}

\bibitem{fido2}
FIDO Alliance, ``FIDO2: Web Authentication Specification,''
W3C Recommendation, 2019. [Online]. Available:
\url{https://www.w3.org/TR/webauthn-2/}

\bibitem{fidouaf}
FIDO Alliance, ``FIDO UAF Architectural Overview,''
FIDO Alliance Specification v1.2, 2017. [Online]. Available:
\url{https://fidoalliance.org/specs/fido-uaf-v1.2-ps-20201012/fido-uaf-overview-v1.2-ps-20201012.html}

\bibitem{fidou2f}
FIDO Alliance, ``FIDO U2F Overview,''
FIDO Alliance Specification v1.2, 2017. [Online]. Available:
\url{https://fidoalliance.org/specs/fido-u2f-v1.2-ps-20170411/fido-u2f-overview-v1.2-ps-20170411.html}

\bibitem{davida98}
G.~Davida, Y.~Frankel, and B.~Matt,
``On enabling secure applications through off-line biometric identification,''
in \emph{Proc.\ IEEE Symp.\ Security Privacy}, 1998, pp.~148--157.

\bibitem{juels99}
A.~Juels and M.~Wattenberg,
``A fuzzy commitment scheme,''
in \emph{Proc.\ ACM Conf.\ Comput.\ Commun.\ Security}, 1999, pp.~28--36.

\bibitem{juels06}
A.~Juels and M.~Sudan,
``A fuzzy vault scheme,''
\emph{Designs, Codes Cryptography}, vol.~38, no.~2, pp.~237--257, 2006.

\bibitem{ratha01}
N.~K.~Ratha, J.~H.~Connell, and R.~M.~Bolle,
``Enhancing security and privacy in biometrics-based authentication systems,''
\emph{IBM Syst.\ J.}, vol.~40, no.~3, pp.~614--634, 2001.

\bibitem{arcface}
J.~Deng, J.~Guo, N.~Xue, and S.~Zafeiriou,
``ArcFace: Additive angular margin loss for deep face recognition,''
in \emph{Proc.\ IEEE/CVF CVPR}, 2019, pp.~4685--4694.

\bibitem{facenet}
F.~Schroff, D.~Kalenichenko, and J.~Philbin,
``FaceNet: A unified embedding for face recognition and clustering,''
in \emph{Proc.\ IEEE/CVF CVPR}, 2015, pp.~815--823.

\bibitem{cucchiara05}
R.~Cucchiara,
``Multimedia surveillance systems,''
in \emph{Proc.\ ACM Workshop Video Surveillance Sensor Netw.}, 2005, pp.~3--10.

\bibitem{jain06}
A.~K.~Jain, A.~Ross, and S.~Pankanti,
``Biometrics: A tool for information security,''
\emph{IEEE Trans.\ Inf.\ Forensics Security}, vol.~1, no.~2,
pp.~125--143, Jun.~2006.

\bibitem{meden2021}
O.~Meden, P.~Peer, and V.~\v{S}truc,
``Privacy-enhancing face biometrics: A comprehensive survey,''
\emph{IEEE Trans.\ Inf.\ Forensics Security}, vol.~16, pp.~4147--4183, 2021.

\bibitem{arcface_journal}
J.~Deng, J.~Guo, N.~Xue, and S.~Zafeiriou,
``ArcFace,''
\emph{IEEE Trans.\ Pattern Anal.\ Mach.\ Intell.}, vol.~44, no.~10,
pp.~5962--5980, 2022.

\bibitem{vggface2}
Q.~Cao, L.~Shen, W.~Xie, O.~M.~Parkhi, and A.~Zisserman,
``VGGFace2: A dataset for recognising faces across pose and age,''
in \emph{Proc.\ IEEE FG}, 2018, pp.~67--74.

\bibitem{lfw}
G.~B.~Huang, M.~Ramesh, T.~Berg, and E.~Learned-Miller,
``Labeled Faces in the Wild: A database for studying face recognition in
unconstrained environments,''
Univ.\ Massachusetts Amherst, Tech.\ Rep.\ 07-49, 2007.

\bibitem{megaface}
I.~Kemelmacher-Shlizerman, S.~M.~Seitz, D.~Miller, and E.~Brossard,
``The MegaFace benchmark: 1 million faces for recognition at scale,''
in \emph{Proc.\ IEEE/CVF CVPR}, 2016, pp.~4873--4882.

\bibitem{facescrub}
H.-W.~Ng and S.~Winkler,
``A data-driven approach to cleaning large face datasets,''
in \emph{Proc.\ IEEE ICIP}, 2014, pp.~343--347.

\bibitem{deepface}
Y.~Taigman, M.~Yang, M.~Ranzato, and L.~Wolf,
``DeepFace: Closing the gap to human-level performance in face verification,''
in \emph{Proc.\ IEEE/CVF CVPR}, 2014, pp.~1701--1708.

\bibitem{deepfr}
O.~M.~Parkhi, A.~Vedaldi, and A.~Zisserman,
``Deep face recognition,''
in \emph{Proc.\ BMVC}, vol.~1, 2015, p.~6.

\bibitem{fidoux}
FIDO Alliance, ``FIDO UX Guidelines,''
FIDO Alliance White Paper, 2017. [Online]. Available:
\url{https://fidoalliance.org/white-paper-fido-ux-guidelines/}

\bibitem{msceleb}
Y.~Guo, L.~Zhang, Y.~Hu, X.~He, and J.~Gao,
``MS-Celeb-1M: A dataset and benchmark for large-scale face recognition,''
in \emph{Proc.\ ECCV}, 2016, pp.~87--102.

\bibitem{webface260m}
Z.~Zhu et~al.,
``WebFace260M: A benchmark for million-scale face recognition,''
in \emph{Proc.\ IEEE/CVF CVPR}, 2021, pp.~10492--10502.

\bibitem{nist800-63b}
P.~A.~Grassi, M.~E.~Garcia, and J.~L.~Fenton,
``NIST Special Publication 800-63B: Digital Identity Guidelines: Authentication
and Lifecycle Management,''
National Institute of Standards and Technology, Gaithersburg, MD, 2017.
[Online]. Available: \url{https://doi.org/10.6028/NIST.SP.800-63b}

\bibitem{mediapipe}
V.~Bazarevsky, Y.~Kartynnik, A.~Vakunov, K.~Raveendran, and M.~Grundmann,
``BlazeFace: Sub-millisecond neural face detection on mobile GPUs,''
\emph{arXiv preprint arXiv:1907.05047}, 2019; see also
I.~Grishchenko, A.~Ablavatski, Y.~Kartynnik, K.~Raveendran, and
M.~Grundmann,
``Attention mesh: High-fidelity face mesh prediction in real-time,''
\emph{arXiv preprint arXiv:2006.10962}, 2020.
[Online]. Available: \url{https://google.github.io/mediapipe/solutions/face_mesh}

\bibitem{nuaa}
X.~Tan, Y.~Li, J.~Liu, and L.~Jiang,
``Face liveness detection from a single image with sparse low rank bilinear
discriminative model,''
in \emph{Proc.\ ECCV}, 2010, pp.~504--517.

\bibitem{casiasurf}
R.~Shao, X.~Lan, J.~Li, and P.~C.~Yuen,
``Multi-adversarial discriminative deep domain generalization for face
presentation attack detection,''
in \emph{Proc.\ IEEE/CVF CVPR}, 2019, pp.~10023--10031.

\bibitem{dlib09}
D.~E.~King,
``Dlib-ml: A machine learning toolkit,''
\emph{J.\ Mach.\ Learn.\ Res.}, vol.~10, pp.~1755--1758, 2009.

\bibitem{opencv}
G.~Bradski,
``The OpenCV library,''
\emph{Dr.\ Dobb's Journal of Software Tools}, vol.~25, no.~11,
pp.~120--125, 2000. [Online]. Available: \url{https://opencv.org}

\bibitem{dpdp2023}
Government of India, ``Digital Personal Data Protection Act, 2023.'' [Online]. Available: \url{https://www.meity.gov.in/static/uploads/2024/06/2bf1f0e9f04e6fb4f8fef35e82c42aa5.pdf}. Accessed: Apr. 15, 2026.

\end{thebibliography}

\end{document}